\documentclass[doublecol,linenumbers,figures]{epl2}
\usepackage{amsthm,amsmath,amsfonts,dsfont,upgreek}    
\usepackage{amssymb,epsfig,setspace}
\usepackage{graphicx}
\usepackage[english]{babel}
\usepackage{verbatim}
\usepackage{bm}

\def\op#1{{\Hat{\mathrm{#1}}}}

\begin{document}
\hyphenation{ge-ne-ra-tes dif-fe-rent}
\hyphenation{me-di-um as-su-ming pri-mi-ti-ve pe-ri-o-di-ci-ty}
\hyphenation{mul-ti-p-le-sca-t-te-ri-ng i-te-ra-ti-ng e-q-ua-ti-on}
\hyphenation{wa-ves di-men-si-o-nal ge-ne-ral the-o-ry sca-t-te-ri-ng}
\hyphenation{di-f-fe-r-ent tra-je-c-to-ries e-le-c-tro-ma-g-ne-tic pho-to-nic}
\hyphenation{Ray-le-i-gh di-n-ger}

\title{Generalized Rabi models: diagonalization in the spin subspace and
differential operators of Dunkl type}

\author{
Alexander Moroz 
} 
\institute{Wave-scattering.com}

\pacs{02.30.Ik}{Integrable systems}
\pacs{42.50.Pq}{Cavity quantum electrodynamics; micromasers}
\pacs{85.75.-d}{Spintronics}

\abstract{
A discrete parity $\mathbb{Z}_2$ symmetry of a two parameter extension of 
the quantum Rabi model which smoothly interpolates between the latter and 
the Jaynes-Cummings model, and of the two-photon and the two-mode quantum Rabi models 
enables their diagonalization in the spin subspace. 
A more general statement is that the respective sets of $2\times 2$ hermitian 
operators of the Fulton-Gouterman type and those diagonal in the spin subspace
are unitary equivalent. The diagonalized
representation makes it transparent that any question about integrability and
solvability can be addressed only at the level of 
ordinary differential operators of Dunkl type.
Braak's definition of integrability is shown (i) to contradict earlier 
numerical studies and (ii) to imply that any physically reasonable 
differential operator of Fulton-Gouterman type is integrable.
}

%

\maketitle

%


\section{Introduction}
\label{sc:intr}
Any $2\times 2$ hermitian operator $\hat{H}$ can be expressed in the form
\begin{equation}
\hat{H}= \sum_{j=0}^3 h_j\sigma_j,\quad
h_j=\tfrac{1}{2}\, \mbox{Tr }(\hat{H}\sigma_j),
\label{gex}
\end{equation}
where $h_j$'s are one-dimensional operators in a suitable Hilbert space $\mathfrak{H}$ 
and here and elsewhere the standard representation of the Pauli matrices 
$\sigma_j$, $j=1,2,3$, is assumed. For the sake of compactness, we set
$\sigma_0:=\mathds{1}$ in summation formulas,
with $\mathds{1}$ being the unit matrix. 
$\hat{H}$ is said to be of the {\em Fulton-Gouterman} type \cite{FG}, 
and denoted by $\hat{H}_{FG}$, if (i) $\hat{H}$ is similar to
\begin{equation}
\hat{H}_{FG}= A\mathds{1} + B\sigma_1 + C\sigma_2+ D\sigma_3,
\label{hfg}
\end{equation}
and (ii) there is a hermitian operator $\op{R}$ such that 
\begin{equation}
[\op{R},A]=[\op{R},B]=0,\quad
\{\op{R},C\}=\{\op{R},D\}=0,
\label{abcd}
\end{equation}
where $[,]$ and $\{,\}$ denote the conventional commutator and anticommutator.
(Our definition of $\hat{H}_{FG}$ is broader than
the original one by including the term $C\sigma_2$ 
({\em cf.} $\hat{H}$ in eq. (4.1) of ref. \cite{FG}).)

A prominent example of the Fulton-Gouterman type Hamiltonians 
will be shown to be the generalized Rabi model (GRM) 
studied by M\"uller {\em et al.} \cite{CCM,SMS},
Schir\'{o} {\em et al.} \cite{SBB}, Gritsev {\em et al.} \cite{TAP,TPG}, and
others \cite{ZZ},
\begin{equation}
 \op{H}_{\mathrm{gR}}
 =
 \gamma a^{\dag}a \!
 +\!
 \mu \sigma_3\!
 +\!
 k_1\! \left( a^{\dag}\sigma_-\! +\! a\sigma_+ \right)
 +
 k_2\! \left( a^{\dag}\sigma_+\! +\! a\sigma_- \right)\!,
 \label{gR}
\end{equation}
and the two-photon (TPRM) and the two-mode quantum Rabi models (TMRM) \cite{Zh2}
[{\em cf.} eq. (\ref{2RM}) below]. Here $\hat{a}$ and $\hat{a}^\dagger$ 
are the conventional boson 
annihilation and creation operators satisfying commutation relation 
$[\hat{a},\hat{a}^{\dagger}] = 1$. In the Fock-Bargmann 
representation \cite{Brg,BG},
\begin{equation}
 a \to {\rm d}/{\rm d}z ={\rm d}_z, \qquad a^{\dagger} \to z,
\label{fbr}
\end{equation}
$\op{H}_{\mathrm{gR}}$ becomes a first-order differential operator
on $\mathfrak{B}\otimes \mathbb{C}^2$, where $\mathfrak{B}$ 
is the Fock-Bargmann Hilbert space of entire analytic functions isomorphic 
to $L^2(\mathbb{R})$ \cite{Brg}. The GRM interpolates between 
the Jaynes-Cummings model (JCM) \cite{JC} (for $k_2=0$) and 
the original quantum Rabi model (RM)
\cite{Rb,Schw,SS,Ks,Schm,Tur1,Br,AMep,AMops,AMef,CR} 
(for $k_1=k_2=k$). The RM describes the simplest interaction between a 
cavity mode with a bare frequency $\omega$ and a two-level system, or a qubit, 
with a bare resonance frequency $\omega_0$.
One has $\gamma=\hbar \omega>0$, $k_1=k_2=\hbar g>0$,
with $g$ being a coupling constant, $\mu=\hbar \omega_0/2$, where
$\hbar$ is the reduced Planck constant.
The RM with a {\em negative} sign of its parameters $g$ and $\mu$ 
({\em cf.} eq. (12) of ref. \cite{SS}) is used to describe an excitation 
hopping between two sites 
($\mu$ is then a tunneling parameter) and is relevant in
understanding the transition between untrapped and trapped behavior of an exciton.
The GRM serves as a non-trivial model in spin
resonance, for various problems involving the interaction between 
electronic and vibrational degrees of freedom in molecules and solids,
and in quantum optics \cite{KGK,BGA,FLM,NDH,CRL}.
The RM and GRM are presently the focus of intense experimental and theoretical activity
for cavity- and circuit-QED setups, superconducting q-bits, nitrogen vacancy centers, 
etc. \cite{TPG,KGK,BGA,FLM,NDH,CRL}. 
With new experiments rapidly approaching the limit of
the so-called deep strong coupling regime, there is every reason to believe that 
such systems could open up a rich vein of research on truly 
quantum effects with implications for quantum information science and 
fundamental quantum optics.
There are several further motivations to consider this model.
It can be mapped onto the model describing
a two-dimensional electron gas with Rashba ($\alpha_{R}\sim k_1$) 
and Dresselhaus ($\alpha_{D}\sim k_2$) spin-orbit couplings subject
to a perpendicular magnetic field (the Zeeman splitting thereby equals $2\mu$)
\cite{TAP,TPG,EES}.
The Rashba spin-orbit coupling (SOC) can be tuned by an applied electric
field while the Zeeman term is tuned by an applied magnetic field. 
This allows us to explore the whole parameter space of
the model. 
In ref. \cite{GS} a possible realization of tunable Rashba and Dresselhaus 
SOC with ultracold alkali atoms is proposed, where each state is coupled by a two-photon
Raman transition.
Further examples of physical realizations of the GRM include 
(i) electric-magnetic coupling of light and matter, and 
(ii) effective realization of the model using $3$- and $4$-level emitters \cite{TPG}.

Provided that $\op{R}^2=1$, any $\hat{H}_{FG}$ enjoys a discrete $\mathbb{Z}_2$-symmetry 
$\hat{\Pi}_{FG}=\op{R}\sigma_1$: $[\hat{H}_{FG},\hat{\Pi}_{FG}]=0$, $\hat{\Pi}_{FG}^2=1$.
The $\mathbb{Z}_2$-symmetry suffices to partially diagonalize
$\hat{H}_{FG}$ operating on the Hilbert space $\mathfrak{B}\otimes \mathbb{C}^2$ 
in the spin subspace \cite{Wg}. Indeed, a sufficient condition
for the spin-subspace diagonalization of a Hamiltonian $\hat{H}$ 
on $\mathfrak{B}\otimes \mathbb{C}^N$ is that $\hat{H}$ possesses an {\em Abelian} 
symmetry $G$ of the order $N$. Surprisingly enough, 
the diagonalization of the GRM, TPRM, and TMRM
in the spin subspace has not been discussed yet - {\em cf.} 
refs. \cite{CCM,SMS,SBB,TAP,TPG,ZZ,Zh2,Rb,Schw,Ks,Tur1,Br,AMep,AMops,AMef,CR} -
even though the diagonalization can be performed by rather straightforward 
unitary transformation: 
\vspace*{0.1cm}

\noindent {\bf Theorem 1}: Any $\hat{H}_{FG}$ given by (\ref{hfg}) 
can be diagonalized in the spin subspace by means of 
a unitary transformation, 
\begin{eqnarray}
U_{FG}\hat{H}_{FG}U_{FG}^{-1} &=& (A+D)\mathds{1} 
+ B\op{R}\sigma_3 - {\rm i}C \op{R}\sigma_3,
\nonumber\\
U_{FG}\hat{\Pi}_{FG}U_{FG}^{-1} &=& \sigma_3,
\label{hfgd}
\end{eqnarray}
induced by 
\begin{equation}
U_{FG} = \frac{1}{2}\,\left[(1+\op{R}) U_{13} + (1-\op{R}) U_2^{-1}
\right],
\label{Ufg}
\end{equation}
where $U_{13}=(\sigma_1 + \sigma_3)/\sqrt{2}$ and
$U_{2}=(\mathds{1} + {\rm i}\sigma_2)/\sqrt{2}$.
Thereby an original spin-$1/2$ problem in the Hilbert space
$\mathfrak{H}\otimes \mathbb{C}^2$ decouples into two
distinct one-dimensional problems in $\mathfrak{H}$, each characterized by 
the operator
\begin{equation}
\op{L}_\pm =A + D \pm (B -{\rm i}C)\op{R},
\label{opL}
\end{equation}
where the $\pm$ sign corresponds to the respective positive 
and negative parity eigenspaces.
\vspace*{0.1cm}

\hfill{$\square$}
\vspace*{0.1cm}

The letter is organized as follows. Theorem 1 is applied to the GRM, 
TPRM and TMRM. For the GRM the operators $\op{L}_\pm$ become 
first-order ordinary differential 
operators of {\em Dunkl} type \cite{Dunkl,ViZd}, 
whereas for the TPRM and TMRM they
become second-order differential operators of Dunkl type.
The Dunkl type operators, which are characterized in that they contain 
a {\em reflection} operator, became a branch of mathematics only as late as 1990 \cite{Dunkl}. 
Working in the diagonalized representation makes it transparent that 
any question about integrability and solvability can be addressed 
only at the level of ordinary differential operators of Dunkl type. 
Braak's definition of integrability \cite{Br} is shown (i) to imply 
that any physically reasonable differential operator of 
Dunkl type is integrable and (ii) to contradict earlier numerical studies 
by M\"uller {\em et al.} \cite{CCM,SMS}.

For completeness an adaption of Wagner's proof directly
to $\hat{H}_{FG}$ as given by (\ref{hfg})
is provided. The reverse of Wagner's theorem is also shown to be true
and the following result is proven:
\vspace*{0.1cm}

\noindent {\bf Theorem 2}:
{\em A hermitian operator $\hat{H}$ has 
the Fulton-Gouterman form (\ref{hfg}), (\ref{abcd})
if and only if $\hat{H}$ is unitary equivalent to an operator 
diagonal in the spin subspace}.
\vspace*{0.1cm}

\hfill{$\square$}


\section{Diagonalization in the spin subspace}
\label{sc:fgtr}
In view of Theorem 1, it suffices to demonstrate the Fulton-Gouterman form
of a given Hamiltonian. $\op{H}_{\mathrm{gR}}$ can be brought
into the Fulton-Gouterman form (\ref{hfg}) upon applying 
a (nonunitary) similarity transformation $\op{\mathcal{H}}_{\mathrm{gR}}=
{\bf W}\op{H}_{\mathrm{gR}}{\bf W}^{-1}$ with
\begin{equation}
 {\bf W} =
 \frac{1}{\sqrt{2}}
\left(
\begin{array}{lr}
 \! w\! &\! w^{-1} \!\\
 \! w\! &\! -w^{-1}\!
 \end{array}
\right),
 \quad
 {\bf W}^{-1}
 =
 \frac{1}{\sqrt{2}} 
\left(
\begin{array}{lc}
 \! w^{-1}\! &\! w^{-1} \!\\
 \! w \! &\! -w\!
\end{array}
\right),
\nonumber 
\end{equation}
where $w=(k_2/k_1)^{1/4}$. Under the similarity transformation
\begin{eqnarray}
&\sigma_+ \to \frac{w^2}{2}\, (\sigma_3-{\rm i}\sigma_2),~~~
\sigma_- \to \frac{1}{2w^2}\, (\sigma_3+{\rm i}\sigma_2),&
\nonumber\\
&
k_1\sigma_+ +k_2\sigma_- \to \sqrt{k_1k_2}\, \sigma_3,~~~
\sigma_3 \to \sigma_1, &
\nonumber\\
&
k_2\sigma_+ +k_1\sigma_- \to \frac{k_1^2+ k_2^2}{\sqrt{k_1k_2}} \sigma_3
+ {\rm i}\frac{k_1^2- k_2^2}{\sqrt{k_1k_2}} \sigma_2.&
\nonumber 
\end{eqnarray}
The transformed $\op{\mathcal{H}}_{\mathrm{gR}}$ divided 
by $\gamma$ takes on the Fulton-Gouterman form (\ref{hfg}) with
\begin{equation}
A=\hat{a}^\dagger \hat{a},~~
B=\Delta,~~~C={\rm i}\frac{\lambda_-}{\kappa}\,\hat{a}^\dagger,~~
D= \kappa a + \frac{\lambda_+}{\kappa}\,\hat{a}^\dagger,
\nonumber
\end{equation}
where 
\begin{equation}
\Delta:=\frac{\mu}{\gamma},~~~
 \lambda_\pm := \frac{k_1^2 \pm k_2^2}{2\gamma^2},~~~
 \kappa :=\frac{\sqrt{k_1k_2}}{\gamma}\cdot
\nonumber 
\end{equation}
The Fulton-Gouterman symmetry operation 
is realized by unitary $\op{R}=e^{{\rm i}\pi \hat{a}^\dagger \hat{a}}$,
which induces {\em reflections} of the annihilation and creation operators:
$\hat{a}\rightarrow-\hat{a}$, $\hat{a}^\dagger\rightarrow-\hat{a}^\dagger$, 
and leaves the boson number operator $\hat{a}^\dagger \hat{a}$ invariant \cite{FG}.
According to Wagner's theorem, $\op{\mathcal{H}}_{\mathrm{gR}}$ 
in the Fock-Bargmann representation (\ref{fbr}) is unitary equivalent to
\begin{equation}
\op{\mathcal{H}}_{\mathrm{gR}} =
 \left[(z + \kappa){\rm d}_z + \frac{\lambda_+z}{\kappa}\right] 
  +\left(\frac{\lambda_-z}{\kappa}+\Delta\right) \sigma_3 \op{R}.
\nonumber 
\end{equation}

In the limit $k_1=k_2=g$ leading to the RM in 
 a unitary equivalent {\em single-mode spin-boson} picture: 
\begin{equation}
\lambda_- = 0, \qquad 
\kappa =\frac{g}{\omega}, \qquad 
 \frac{\lambda_+}{\kappa} =\kappa,
\nonumber 
\end{equation}
\begin{equation}
A= z{\rm d}_z,~~~
B= \Delta,~~~
C=0,~~~ D=\kappa (z + {\rm d}_z),
\nonumber 
\end{equation}
(we set the reduced Planck constant $\hbar=1$) 
and ${\bf W}$ becomes the unitary transformation $U_{13}$ \cite{Ks,AMops}:
\begin{equation}
{\bf W} =
 \frac{1}{\sqrt{2}}
\left(
 \begin{array}{lr}
 \! 1\! &\! 1\! \\
 \! 1\! &\! - 1\!
 \end{array}
\right) =\frac{1}{\sqrt{2}}\, (\sigma_1+\sigma_3)\equiv U_{13}=
 U_{13}^{-1}.
\nonumber 
\end{equation}
(A different ${\bf W}$ has been used by Gritsev {\em et al.} \cite{TPG} 
which in its symmetrized form reduces to $(\sigma_1-\sigma_3)/\sqrt{2}$
in the limit $k_1=k_2$.) 

The Hamiltonians 
${\op{H}_{2p}\!=\!\omega a^\dagger a+\beta\sigma_3+g\,\sigma_1\!\left[(a^\dagger)^2+a^2\right]}$ 
and ${\op{H}_{2m}=\omega(a_1^\dagger a_1+a_2^\dagger a_2)+\beta\sigma_3+
g\,\sigma_1(a_1^\dagger a_2^\dagger+a_1 a_2)}$ of the TPRM and the TMRM, 
respectively, are on unitary transforming with $U_{13}$ brought into the Foulton-Gouterman form
({\em cf.} eqs. (4.3) and (5.3) of \cite{Zh2})
\begin{equation}
\op{H}_{2FG}=\gamma \left(K_0-c \right)+\Delta\sigma_1+\sigma_3(K_++K_-),
\label{2RM}
\end{equation}
where $\gamma=\omega/g$, $\Delta=\beta/g$.
Compared to the {\em Heisenberg} algebra of $a$, $a^\dagger$ in $\op{H}_{\mathrm{gR}}$, 
the operators $K_\pm, K_0$ in (\ref{2RM}) 
form the usual $su(1,1)$ Lie algebra, $[K_0, K_\pm]=\pm K_\pm$, $[K_+, K_-]=-2K_0$.
In the case of the TPRM, $c=1/4$,
\begin{equation}
K_+=\tfrac{1}{2}(a^\dagger)^2,~~~K_-=\tfrac{1}{2}a^2,~~~
K_0=\tfrac{1}{2}\left(a^\dagger a+\tfrac{1}{2}\right),
\nonumber 
\end{equation}
whereas, in the case of the TMRM, $c=1/2$, and
\begin{equation}
K_+=a_1^\dagger a_2^\dagger,~~~K_-=a_1 a_2,~~~
K_0=\tfrac{1}{2}(a_1^\dagger a_1+a_2^\dagger a_2+1).
\nonumber 
\end{equation}
The parity symmetry $\hat{\Pi}_{FG}=\op{R}\sigma_1$ is 
realized by unitary $\op{R}=e^{{\rm i}\pi K_0}$, which
induces reflections of $K_\pm$ and leaves $K_0$ invariant.

The Fock-Bargmann Hilbert space $\mathfrak{B}$ is based on the coherent states
associated with the Heisenberg algebra \cite{BG}.
In the present case, $\mathfrak{B}$ gets replaced by a 
more general Hilbert space of entire analytic functions of growth $(1,1)$ associated with 
the so-called Barut-Girardello coherent states \cite{BG} 
of the annihilation operator $K_-$ of the $su(1,1)$ Lie algebra.
In an infinite-dimensional unitary irreducible representation, 
known as the positive discrete series ${\cal D}^+(q)$,
the operators $K_\pm, K_0$ of the TPRM, which realize the single-mode bosonic
representation of $su(1,1)$, are represented as differential operators,
\begin{equation}
K_0 = z{\rm d}_z+ q, ~~~K_+ = z/2, ~~~
K_- =2z{\rm d}_z^2 +4q{\rm d}_z,
\nonumber 
\end{equation}
where the parameter $q$, called Bargmann index, sati\-sfies $q=1/4, 3/4$.
The operators $K_\pm, K_0$ of the TMRM providing the two-mode bosonic
representation of $su(1,1)$ have the single-variable differential realization as
\begin{equation}
K_0=z{\rm d}_z+q,~~~K_+=z,~~~~K_-=z{\rm d}_z^2 +2q{\rm d}_z,
\nonumber 
\end{equation}
where $q>0$ can be any integer or half-integer \cite{Zh2,BG}.

\section{Linear differential operators of Dunkl type}
The action of $\op{R}$ in the above cases 
reduces to reflections $\op{R}f(z)=f(-z)$ in 
a suitable Hilbert space of entire analytic functions.
In the case of the GRM, the respective diagonal components 
$\op{L}_\pm$ defined by (\ref{opL})
become linear first-order ordinary 
differential operators of Dunkl type,
\begin{equation}
\op{L}_\pm = (z+\kappa){\rm d}_z + \frac{\lambda_+z}{\kappa} 
          \pm \left(\frac{\lambda_-z}{\kappa}+\Delta\right) \op{R}.
\label{opLr}
\end{equation}
In the limit of the RM one recovers
\begin{equation}
\op{L}_\pm =(z+\kappa) {\rm d}_z + \kappa z \pm \Delta \op{R}
\label{opR}
\end{equation}
({\em cf.} eq. (21) of ref. \cite{SS} and eq. (2.1) of ref. \cite{Schm}).
For the respective TPRM and the TMRM one finds
\begin{eqnarray}
&\op{L}_{\pm;2p} = 2z{\rm d}_z^2+(4q+ \gamma z) {\rm d}_z 
      + \tfrac{z}{2}+\gamma \left(q -\tfrac{1}{4}\right) \pm \Delta \op{R},\quad&
\nonumber  
\\
&\op{L}_{\pm;2m} = z{\rm d}_z^2+(2q+ \gamma z) {\rm d}_z 
      + z+\gamma \left(q -\tfrac{1}{2}\right) \pm \Delta \op{R}.\quad&
\label{opL2m}
\end{eqnarray}
For a general $\op{L}_\pm$ in (\ref{opL}) one has
\begin{equation}
[\op{L}_\pm,\op{R}]=\mp\, 2{\rm i}C+2D\op{R} \ne 0.
\nonumber 
\end{equation}
Therefore also $[\op{L}_\pm,\op{R}\op{L}_\pm]\ne 0$.
Whereas for an eigenvector $\phi$ of $\op{L}_\pm$ 
one has $\op{R}\op{L}_\pm \phi =\epsilon\op{R}\phi$, 
one cannot say anything definite about $\op{L}_\pm\op{R} \phi$.

Note that in the absence of the reflection operator 
$\op{R}$ in eq. (\ref{opR}), {\em e.g.} with 
$\op{L}_\pm =(z+\kappa) {\rm d}_z + \kappa z \pm \Delta$,
the eigenvalue problem,
\begin{equation}
(\op{L}-\epsilon)\phi=0,
\label{leig}
\end{equation}
can be easily integrated. One finds 
$\phi(z)= \mbox{const} (z+\kappa)^{\kappa^2\pm \Delta -\epsilon}{\rm e}^{-\kappa z}$,
where `const' is an integration constant.
The solutions will be holomorphic if and only if
$\kappa^2\pm \Delta -\epsilon \in\mathbb{N}_0$.
In spite of a deceptive simplicity of $\op{L}_\pm$ in eqs. (\ref{opLr}), (\ref{opR}), 
a rigorous analytic solution of (\ref{leig}) remains an unsolved problem
(in the sense that analytic expressions for eigenvalues are not known - {\em cf.} ref. \cite{Mnn}). 
This demonstrates that the reflection operator $\op{R}$ is a highly nontrivial
obstruction for solving the eigenvalue problem (\ref{leig}) \cite{ViZd}.

Each term of the one-dimensional operator $\op{L}_\pm$ in (\ref{opLr}), (\ref{opR}), (\ref{opL2m}) 
does not change the degree of a monomial $z^n$ by more than $\pm 1$.
Thereby the resulting eigenvalue problem 
(\ref{leig}) naturally reduces to a {\em three-term recurrence relation} (TTRR) 
\cite{Zh2,Schw,Tur1,AMep,AMops,AMhd}. Thus each $\op{L}_\pm$ corresponds to 
the so-called irreducible component of ref. \cite{AMef}. Those have been shown 
to have a {\em nondegenerate} spectrum under very general conditions, and hence no level crossing while 
varying coupling parameter(s). Alternatively, the nondegeneracy applies to all problems 
where the Hamiltonian operator is a self-adjoint extension of a {\em tridiagonal} Jacobi matrix 
of deficiency index $(1,1)$ \cite{Tur1,FuH}. Therefore, all $\hat{H}_{FG}$ leading to a TTRR
have {\em avoided} level crossings. 
The above arguments provide rigorous proof for the avoided crossing 
observed numerically by M\"uller {\em et al.} \cite{CCM,SMS}.

\section{Braak's definition of integrability}
According to Braak's definition of integrability \cite{Br}:
If each eigenstate of a quantum system with $f_1$ discrete and $f_2$
continuous degrees of freedom can be {\em uniquely} labeled by
$f_1 + f_2 = f$ quantum numbers $\{d_1, \ldots, d_{f_1},c_1, \ldots, c_{f_2}\}$, 
such that the $d_j$ can take on dim$(H_j)$ different values, where $H_j$ is the state
space of the $j$th discrete degree of freedom and the $c_k$ range from
$0$ to infinity, then this system is {\em quantum integrable}.
The RM has $f_1=f_2=1$ and degeneracies take place
only between levels of states with {\em different} parity, whereas within the
parity subspaces no level crossings occur. The global label for the RM
[valid for all values of $\kappa$ in eq. (\ref{opR})] is two dimensional,
with one label for the parity and the other being the energy sorting number
within a given parity subspace, and hence, according to Braak, 
the RM is quantum {\em integrable}. But this leads to an inflation of 
integrable models, because the {\em avoided} level crossings between states of equal 
parity is generic for the models
studied here. Following Braak's arguments, all physically reasonable 
$\hat{H}_{FG}$ are necessarily quantum integrable.

Braak's arguments appear to be based on a wrong assessment of the role of {\em discrete} symmetries.
The latter divide the Hilbert space of $\hat{H}_{FG}$ into invariant subspaces. In
general, this does not result in symmetry-induced level-degeneracies, but it
does lead to accidental degeneracies between levels belonging to different
invariant subspaces ({\em cf.} Judd solutions). Such level crossings exist 
independently of whether or not $\hat{H}_{FG}$ is integrable \cite{SMS}. 

The general rule has always been to analyze corresponding invariant 
or irreducible components. 
For instance, in statistical analysis of the eigenvalues of quantum billiards 
one performs the so-called {\em desymmetrization},
which reduces the study to a fundamental domain of a discrete group \cite{BGS}.
The diagonalized representation makes it transparent that integrability and
solvability of the GRM can be addressed only at the level of 
first-order one-dimensional differential operators $\op{L}_\pm$.
The latter necessitates considering each invariant parity subspace independently.
In their thorough numerical studies, M\"uller {\em et al.} \cite{CCM,SMS} did just that.
They made use of the fact that a quantum integrability cannot be inferred from {\em quantum invariants} as simply 
as classical integrability can be inferred from integrals of the motion ({\em analytic invariants}). 
Commuting operators can always be constructed {\em irrespective} of whether the model is (classically)
integrable or not \cite{CCM,SMS,Per}. When Einstein-Brillouin-Keller quantization 
is possible, it applies to all conserved dynamical variables (not only to the Hamiltonian) 
and in particular to the {\em time average} of {\em any} dynamical variable. 
Any operator $T$ that is not already an invariant, $[H,T]\neq 0$, 
can be turned into an invariant via time average. 
In the {\em energy} representation, the time average strips $T$ of all 
its off-diagonal elements. The resulting operator 
$I_T =\langle T \rangle$ being {\em diagonal} in the energy representation 
thus commutes with $H$ by construction \cite{Per}. 
M\"uller {\em et al.} \cite{CCM,SMS} studied two-dimensional patterns 
of quantum invariants $\{(\epsilon_n,\langle T_j \rangle_n)\}$, 
where $\epsilon_n$ is the $n$th eigenvalue, and
$T_1 = a^\dagger\sigma_+$, $T_2 = a^\dagger(\sigma_- + \sigma_+)$.
(Although $T_j$'s are not hermitian, the matrix elements 
$\langle n |T_j| n \rangle= \langle T_j \rangle_n$
happen to be real for all energy eigenstates \cite{CCM}.)
The patterns of points $\{(\epsilon_n,\langle T_j \rangle_n)\}$
in invariant parity subspaces were found to 
be strikingly different for the respective integrable ($k_1k_2=0$) 
and nonintegrable ($k_1k_2\ne 0$) cases. 
A {\em qualitative} change in pattern required the assignment of 
mutually exclusive sets of quantum numbers 
to the same set of eigenstates in different parameter regimes \cite{CCM,SMS}.
In the integrable cases, the patterns formed two separate linear strands of points.
Level crossings required a two dimensional label for an unambiguous 
assignment of levels, each label corresponding to one of the respective quantum invariants.
Contrary to that, a single {\em level sorting} quantum number sufficed to 
label all eigenstates in the presence of the {\em avoided} level crossings 
between states of equal parity for nonintegrable cases. In contrast to Braak's conclusion,
avoided level crossings were found to be the trademark of quantum {\em nonintegrability}. 
The integrable and nonintegrable cases revealed 
also unambiguously different patterns of coordinated motion of all states 
with given parity in the plane of invariants $(\epsilon_n,\langle T_2\rangle_n)$ 
as the interaction strength ({\em i.e.} $\Lambda$ in the parametrization 
$k_1=\Lambda\cos\alpha$, $k_2=\Lambda\sin\alpha$ of the coupling constants) 
gradually increased \cite{SMS}.
The distinctive attributes of quantum invariants in the integrable and
nonintegrable regimes of a quantum system are subtle but {\em not}
ambiguous. As soon as $k_1k_2\ne 0$ (or $\alpha\ne 0,\pi/2$), 
the GRM was found {\em nonintegrable} \cite{CCM,SMS}.

\section{Proof of Theorem 1}
\label{sec:diag}
It is expedient to introduce 
\begin{equation}
U_{jkl}= \frac{1}{2}\,\left[(1+\op{R}) U_{jk} + (1-\op{R}) U_{l}
\right]
\label{Ujkl}
\end{equation}
with unequal $j,k,l=1,2,3$, where
\begin{equation}
U_{jk}=(\sigma_j + \sigma_k)/\sqrt{2},\quad
U_{l}=(\mathds{1} + {\rm i}\sigma_l)/\sqrt{2},
\label{mbt}
\end{equation}
are $2\times 2$ unitary matrices.
Any $U_{jkl}$ is thus a linear combination of unitary matrices with the coefficients
being one-dimensional projectors $\op{P}_\pm=(1\pm\op{R})/2$.
$U_{jkl}$ itself is unitary:
\begin{eqnarray}
U_{jkl}U_{jkl}^{-1} &=&  \frac{1}{8}\,\left[
(1+\op{R}) U_{jk}+(1-\op{R}) U_{l}
\right]
\nonumber\\
&&
\qquad\qquad \times \left[
(1+\op{R}) U_{jk}+
(1-\op{R})U_{l}^{-1}
\right]
\nonumber\\
&=& 
\frac{1}{4}\,\left[(1+\op{R})^2 +(1-\op{R})^2 \right]\mathds{1}
\nonumber\\
&=&
\frac{1}{4}\,\left[2+2\op{R} +2-2\op{R}\right]\mathds{1}
=
\mathds{1}.
\nonumber 
\end{eqnarray}
Now,
\begin{eqnarray}
\lefteqn{
U_{jkl} \sigma_j U_{jkl}^{-1}
}
\nonumber\\
 &&= \frac{1}{4}
\left[(1+\op{R})^2 U_{jk}\sigma_jU_{jk} 
       + (1-\op{R})^2 U_{l}\sigma_jU_{l}^{-1}
\right]\!.
\nonumber  
\end{eqnarray}
For unequal $j,k,l=1,2,3$ one finds 
\begin{eqnarray}
&U_{jk}\sigma_j U_{jk} = \sigma_k,\quad U_{jk}\sigma_l=- \sigma_l U_{jk},
              \quad U_{l}\sigma_j=\sigma_j U_{l}^{-1},&
\nonumber\\
&U_l\sigma_jU_l^{-1}=-\epsilon_{ljk}\sigma_k,\quad U_l^{-1}\sigma_jU_l=\epsilon_{ljk}\sigma_k,&
\label{mbtr}
\end{eqnarray}
where $\epsilon_{ljk}$ is the usual totally antisymmetric Levi-Civita symbol.
In arriving at the final results we have repeatedly used
\begin{equation}
-{\rm i}\sigma_1\sigma_2\sigma_3=\mathds{1},\quad 
      \sigma_j\sigma_k={\rm i}\epsilon_{jkl} \sigma_l.
\label{pi}
\end{equation}
Under the action of unitary $U_l$ the matrix $\sigma_l$ remains invariant.
It holds trivially that $U_{jkl}\sigma_0 U_{jkl}^{-1} =\sigma_0$.
Given the properties (\ref{mbtr}), one can verify that 
(modulo a sign change and multiplication by $\op{R}$):
\begin{itemize}

\item[($\mathbb{\ast}$)]
the unitary transformation induced by $U_{jkl}$ with unequal $j,k,l=1,2,3$ 
interchanges $\sigma_j$ and $\sigma_k$ while leaving 
$\sigma_0$ and $\sigma_l$ invariant.

\end{itemize}
For any operator $\op{X}$ on $\mathfrak{H}$ commuting with $\op{R}$
\begin{equation}
U_{jkl} \op{X}\sigma_j U_{jkl}^{-1} = \op{X} (U_{jkl} \sigma_j U_{jkl}^{-1}),
\label{Xtr}
\end{equation}
whereas for any operator $\op{Y}$ on $\mathfrak{H}$ anticommuting with $\op{R}$ 
\begin{equation}
U_{jkl} \op{Y}\sigma_j U_{jkl}^{-1} =
     (U_{jkl} \op{Y}U_{jkl}^{-1}) (U_{jkl}\sigma_j U_{jkl}^{-1}).
\label{Ytr}
\end{equation}

\vspace*{0.1cm}

\noindent {\bf Table 1.} $U_{FG}= \frac{1}{2}\,[(1+\op{R}) 
            U_{13} + (1-\op{R}) U_2^{-1}]$.
\begin{center} 
\begin{tabular}{|c|c|c|c|c|} 
\hline 
$\op{O}$ & \quad $A\mathds{1}$ \quad & \quad $B\sigma_1$\quad
               & \quad $C\sigma_2$\quad & \quad $D\sigma_3$ \quad 
\\
\hline 
$U_{FG}\op{O}U_{FG}^{-1}$ & \quad $A\mathds{1}$\quad & $B\op{R} \sigma_3$ & 
            $- {\rm i}C \op{R}\sigma_3$ & $D\mathds{1}$ 
\\
\hline 
\end{tabular}
\end{center}
\vspace*{0.1cm} 
%
The unitary transformation $U_{FG}$ of Theorem 1 
is a particular case of $U_{jkl}$ defined by (\ref{Ujkl}).
One has $U_{FG}=U_{132}^{-1}=U_{13,-2}$, where the minus sign 
in front of $2$ stands for the inverse of $U_2$ in the definition (\ref{Ufg}).
On combining relations (\ref{mbtr}),
\begin{equation}
U_{FG}\sigma_1 U_{FG}^{-1} = \frac{1}{4}\, \sigma_3[(1+\op{R})^2-(1-\op{R})^2]
=\op{R} \sigma_3.
\nonumber  
\end{equation}
Hence for $A$ and $B$ commuting with $\op{R}$ one has
$U_{FG} A\sigma_0 U_{FG}^{-1}=A \sigma_0$ and 
$U_{FG} B\sigma_1 U_{FG}^{-1}=B\op{R} \sigma_3$, respectively.
With the help of identities (\ref{mbtr}) one has
\begin{eqnarray}
&U_{FG} \sigma_0 U_{FG}^{-1}=\sigma_0,\quad
U_{FG} \sigma_1 U_{FG}^{-1}=\op{R} \sigma_3,&
\nonumber\\
&U_{FG} \sigma_2 U_{FG}^{-1}=- \op{R} \sigma_2,\quad
U_{FG} \sigma_3 U_{FG}^{-1}=\sigma_1,&
\label{Uftr}
\end{eqnarray}
conforming to the general rule ($\mathbb{\ast}$).
Because 
\begin{equation}
U_{jk}U_l = U_l^{-1}U_{jk} =
\frac{1}{2}[\sigma_j+\sigma_k-\epsilon_{jkl}(\sigma_j-\sigma_k)],
\nonumber 
\end{equation}
{\em i.e.}, $U_{13}U_2= U_2^{-1} U_{13}=\sigma_1$, one has 
\begin{eqnarray}
\lefteqn{
U_{FG} \op{Y}U_{FG}^{-1} = 
\frac{1}{4} \op{Y} \left[(1-\op{R})^2 U_{13}U_{2} + 
(1+\op{R})^2 U_{2}^{-1}U_{13}\right]
}
\nonumber\\
 && = \frac{1}{4} \op{Y} \sigma_1
\left[(1-\op{R})^2 + 
(1+\op{R})^2 \right] = \op{Y}\sigma_1.\hspace{1cm}
\label{Yfg}
\end{eqnarray}
Eventually, on combining (\ref{Ytr}), (\ref{Uftr}), and (\ref{Yfg}):
\begin{eqnarray}
 U_{FG}C\sigma_2 U_{FG}^{-1} &=&
C \sigma_1 U_{FG}\sigma_2 U_{FG}^{-1}
= - {\rm i}C \op{R}\sigma_3,
\nonumber  
\\
U_{FG}D\sigma_3 U_{FG}^{-1} &=&
D \sigma_1 U_{FG}\sigma_3 U_{FG}^{-1}
= D\mathds{1}.
\label{hfgds3}
\end{eqnarray}
Therefore, the action of $U_{FG}$ summarized in Table 1 ensures 
that any $\hat{H}_{FG}$ of the Fulton-Gouterman type defined by (\ref{hfg}), (\ref{abcd})
can indeed be diagonalized in the spin subspace. 
The form of unitary transformed $\hat{\Pi}_{FG}$ and of operators 
$\op{L}_\pm$ in (\ref{opL}) can be read off from Table 1. 
Thereby the proof is completed.

\section{Proof of Theorem 2}
\label{sc:fgt}
If a hermitian operator $\hat{H}$ has the Fulton-Gouterman form 
(\ref{hfg}), (\ref{abcd}), then, according to Theorem 1, it is unitary 
equivalent to an operator diagonal in the spin subspace. 
Hence in order to prove Theorem 2 it suffices
to show that the reverse holds, too. 

A hermitian operator $\hat{H}$ is diagonal in the spin subspace
if and only if $h_1=h_2\equiv 0$ in the expansion (\ref{gex}).
Now any $h_j\ne 0$ in (\ref{gex}) can be decomposed as 
$h_j=\op{X}_j+\op{Y}_j$, where $[\op{X}_j,\op{R}]=0$ and $\{\op{Y}_j,\op{R}\}=0$,
with $\op{R}$ being an arbitrary reflection operator. To this end, one takes
\begin{equation}
\op{X}_j=\frac{1}{2}(h_j +\op{R}h_j\op{R}),\quad
\op{Y}_j=\frac{1}{2}(h_j -\op{R}h_j\op{R}).
\nonumber 
\end{equation}
A unitary $U$ which commutes with any $\op{X}_j$, $j=0,3$, and brings 
a diagonal operator $\hat{H}$ into
the Fulton-Gouterman form has to necessarily satisfy 
\begin{equation}
U\sigma_3 U^{-1}=\sigma_1.
\label{t2cd}
\end{equation}
At the same time, the transformed set 
$\{U\op{Y}_0\sigma_0 U^{-1},U\op{Y}_3\sigma_3 U^{-1}\}$
has to become $\{\op{Y}'\sigma_2,\op{Y}''\sigma_3\}$,
where the set $\{\op{Y}',\, \op{Y}''\}$ is, modulo a possible sign change and multiplication 
by $\op{R}$ and a constant, equivalent to $\{\op{Y}_0,\op{Y}_3\}$.
In conformity to the general rule ($\mathbb{\ast}$), the condition (\ref{t2cd}) fixes
$U_{jkl}$ to be either $U_{FG}=U_{132}^{-1}$ or $U_{132}$.
The first choice can be excluded in virtue of the second of eqs. (\ref{hfgds3}).
In the case of $U_{132}$, one finds with the help of identities (\ref{mbtr}) 
\begin{eqnarray}
&U_{132} \sigma_0 U_{132}^{-1}=\sigma_0,\quad
U_{132} \sigma_1 U_{132}^{-1}= \sigma_3,&
\nonumber\\
&U_{132} \sigma_2 U_{132}^{-1}= -\op{R} \sigma_2,\quad
U_{132} \sigma_3 U_{132}^{-1}= \op{R}\sigma_1,&
\label{Utr}
\end{eqnarray}
which is consistent with eqs. (\ref{Uftr}).
Because $U_{jk}$ is symmetric in its indices, one can always 
adopt the convention that, when calculating the products 
$U_{jk}U_l^{-1} = U_lU_{jk}$ with unequal $j, k, l$, 
the indices are ordered such that $\epsilon_{jkl}=1$.
With the above convention
\begin{equation}
U_{jk}U_l^{-1} = U_lU_{jk} = \sigma_j,
\nonumber 
\end{equation}
{\em i.e.}, $U_{31}U_{2}^{-1}=U_{2}U_{31}=\sigma_3$,
and one finds [{\em cf.} eq. (\ref{Yfg})]
\begin{equation}
U_{132}\op{Y}U_{132}^{-1}=\op{Y}\sigma_3.
\label{Y3}
\end{equation}
Eventually, in virtue of identities (\ref{Ytr}), (\ref{Utr}), (\ref{Y3}),
\begin{eqnarray}
&U_{132} \op{Y}\sigma_0 U_{132}^{-1}=\op{Y}\sigma_3,\quad
U_{132} \op{Y}\sigma_1 U_{132}^{-1}= \op{Y}\mathds{1},&
\nonumber\\
&U_{132} \op{Y}\sigma_2 U_{132}^{-1}={\rm i}\op{Y}\op{R} \sigma_1,\quad
U_{132} \op{Y}\sigma_3 U_{132}^{-1}=i \op{Y}\op{R} \sigma_2.&
\nonumber  
\end{eqnarray}
Therefore, the unitary transformation induced by $U_{132}$ transforms
the set $\{(\op{X}_0+\op{Y}_0)\sigma_0, (\op{X}_3+\op{Y}_3)\sigma_3\}$
into $\{\op{X}_0\sigma_0,\op{X}_3\op{R}\sigma_1, 
  {\rm i} \op{Y}_3\op{R}\sigma_2, \op{Y}_0\sigma_3\}$,
thereby yielding the Fulton-Gouterman form (\ref{hfg}), (\ref{abcd}). 
The proof is completed.

\section{Conclusions}
\label{sec:conc}
The respective sets of $2\times 2$ hermitian operators of the 
Fulton-Gouterman type and those diagonal in the spin subspace
were shown to be unitary equivalent. As an example,
discrete parity $\mathbb{Z}_2$ symmetry of a two parameter extension of 
the Rabi model which smoothly interpolates between the latter and 
the Jaynes-Cummings model, the so-called generalized Rabi model (GRM),
and of the two-photon and the two-mode quantum Rabi models 
was shown to enable their diagonalization in the spin subspace.
The demonstrated diagonalized representation is expected to greatly simplify 
the description of time evolution and dissipative dynamics of the models. 
In the case of the GRM, supersymmetry on certain submanifolds in a parameter 
space has been established by Gritsev {\em et al.} \cite{TPG}.
The diagonalized representation could facilitate here a much straightforward
identification of supercharges by halving the dimensions of matrices involved.

An intimate relation of the generalized Rabi models with the class of 
differential operators of Dunkl type was established. 
Hopefully, this will help to address computational issues more efficiently.
Many problems involving parity symmetry appear as potential candidates 
of further examples where one could encounter the Dunkl type operators. 
The diagonalization can be straightforwardly extended to spin $s>1/2$ models 
which possess an Abelian symmetry of the order of $N=2s+1$ \cite{Wg,ZhN}.
However the relation with the Dunkl type operators seems to be particular 
for spin $s=1/2$ models: for $N>2$ the Dunkl type operators are associated, in general, 
to nonabelian Coxeter groups \cite{Dunkl}. 
 
The well known level-statistics criteria which have been applied with 
great success to autonomous particle systems
are not applicable to the generalized Rabi models.
The nearest-neighbour distribution of levels is not of the general 
type associated with chaotic systems and does not 
offer any conclusive evidence for quantum nonintegrability \cite{Ksc}.
Only the analysis of two-dimensional patterns of quantum invariants 
$\{(\epsilon_n,\langle T\rangle_n)\}$ yields an unambiguous answer here. 
Braak's definition of integrability was shown not only to contradict 
the earlier pattern studies by M\"uller {\em et al.} \cite{CCM,SMS} 
but also to imply that any physically reasonable 
differential operator of Fulton-Gouterman type ({\em i.e.} leading to a TTRR) is integrable.
This suggests that Braak's definition of integrability is most probably a faulty one.
This is supported by the conclusions of ref. \cite{BaZ} that the Rabi model is 
{\em not} Yang-Baxter integrable.

\acknowledgments

Continuous support of MAKM and discussions with B.~M.~Rodr\'{\i}guez-Lara 
and A. Zhedanov are gratefully acknowledged.


\end{document}